\def\year{2020}\relax
\documentclass[letterpaper]{article} 
\usepackage{aaai20}  
\usepackage{times}  
\usepackage{helvet} 
\usepackage{courier}  
\usepackage[hyphens]{url}  
\usepackage{graphicx} 
\usepackage{graphicx}  
\usepackage{subfigure}
 \usepackage{algorithm}
\usepackage{algorithmic}
\usepackage{booktabs} 
\usepackage{amsmath}
 \usepackage{multirow}
 \usepackage{rotating}
 \usepackage{amssymb}
 \usepackage{bbding}
 \usepackage{amsthm}
 \usepackage{bm}
 \usepackage{epstopdf}
 \usepackage{enumerate}
 \usepackage{array}
 \usepackage{mathtools}
\title{Fast Adaptively Weighted Matrix Factorization for Recommendation with Implicit Feedback}
\author{Jiawei Chen\textsuperscript{\rm{1,2}}, Can Wang\textsuperscript{\rm{1,2,}}\thanks{Corresponding author: wcan@zju.edu.cn}, Sheng Zhou\textsuperscript{\rm{1}}, Qihao Shi\textsuperscript{\rm{1}}, \\ \Large \textbf{Jingbang Chen\textsuperscript{\rm{1}}, Yan Feng\textsuperscript{\rm{1,2}}, Chun Chen\textsuperscript{\rm{1}}} \\
\textsuperscript{\rm{1}} College of Computer Science, Zhejiang University, China\\ \textsuperscript{\rm{2}} Zhejiang University-LianlianPay Joint Research Center \\
\{sleepyhunt,wcan,zhousheng$\_$zju,shiqihao321,chenjb,fengyan,chenc\}@zju.edu.cn
}

\begin{document}

\maketitle

\begin{abstract}
Recommendation from implicit feedback is a highly challenging task due to the lack of the reliable observed negative data. A popular and effective approach for implicit recommendation is to treat unobserved data as negative but downweight their confidence. Naturally, how to assign confidence weights and how to handle the large number of the unobserved data are two key problems for implicit recommendation models. However, existing methods either pursuit fast learning by manually assigning simple confidence weights, which lacks flexibility and may create empirical bias in evaluating user's preference; or adaptively infer personalized confidence weights but suffer from low efficiency.

To achieve both adaptive weights assignment and efficient model learning, we propose a fast adaptively weighted matrix factorization (FAWMF) based on variational auto-encoder. The personalized data confidence weights are adaptively assigned with a parameterized neural network (function) and the network can be inferred from the data. Further, to support fast and stable learning of FAWMF, a new specific batch-based learning algorithm fBGD has been developed, which trains on all feedback data but its complexity is linear to the number of observed data. Extensive experiments on real-world datasets demonstrate the superiority of the proposed FAWMF and its learning algorithm fBGD.
\end{abstract}

\section{Introduction}
 \allowdisplaybreaks[4]
Recommender systems play an important role in many internet services. Since in practise most of recommender systems only have the implicit feedback (e.g. items consumed), recently research attention is increasingly shifted from explicit feedback (e.g. rating prediction) to implicit feedback. However, learning a recommender system from implicit feedback is more challenging. In implicit feedback scenarios, only positive feedback are observed, and the unobserved user-item feedback data (e.g. a user has not bought an item yet) are a mixture of real negative feedback (i.e. a user does not like it) and missing values (i.e. a user just does not know it). Existing methods address this problem by treating all the un-observed data as negative (dislike) but downweighting their confidence. Although it is reasonable, it poses two important questions: (1) How to assign confidence weights for each data? (2) How to handle the massive volume of the unobserved data efficiently?

While there is a rich literature on recommendation from implicit feedback, to our knowledge, all existing methods lack one or more desiderata.  Most of these methods rely on the meticulous assignment of confidence weights to the data. For example, WMF \cite{hu2008collaborative} and eALS \cite{he2016fast} give uniform or popularity-based confidence weights; BPR \cite{rendle2009bpr}, CDAE \cite{wu2016collaborative}, or other sophisticated models downweight the contributions of the negative data in their heuristical (e.g. uniformly) negative sampling strategy. However, choosing these weights usually involves heuristic alterations to the data and needs expensive exhaustive grid search via cross-validation. Furthermore, it is unrealistic for researchers to manually set flexible and diverse weights for millions of data. In practical scenarios, the data confidence weights may change for various user-item combinations. Some unobserved data can be attributed to user's preference while others are the results of users' limited scopes. To this end, exposure-based matrix factorization (EXMF) \cite{liang2016modeling} has been proposed to downweight negative data automatically by predicting how a user knows a item. However, EXMF involves iterative inference of the exposure for each data, which is computationally expensive and also potentially suffers from overfitting. More recently, SamWalker \cite{chen2019samwalker} adaptively assign confidence weights from users' social contexts. However, the social network information is hard obtained in many recommender systems.

To address these problems, we propose a fast adaptively weighted matrix factorization (FAWMF) based on variational auto-encoder \cite{kingma2013auto} for both adaptive weights assignment and efficient learning. We first analyze EXMF from variational perspective and find that the variational posterior of user's exposure acts as data confidence weights in learning users' preference with MF. It is consistent with our intuitions. Only if the user is exposed to the item, he can decide whether to consume the item based on his preference. The data with larger exposure are more reliable in deriving user's preference. Further, we propose to replace confidence weights (variational posterior of user's exposure) with a parameterized inference neural network (function), which reduces the number of the inferred parameters and is capable of capturing latent correlations between users' exposure. In fact, the independent assumption of the exposure is not practical in real world. Typically, recent literatures \cite{palla2005uncovering,zhou2011understanding} in social science claim that each of us belongs to some information-sharing communities. Naturally, users are exposed to these communities and thus their exposure exhibit correlations when they belong to common communities. Motivated by this point, a specific community-based inference neural network has been designed, which explores latent communities among users and infer users' exposure from the communities that they belong to. It will capture more precise exposure and can mitigate the over-fitting problem. By optimizing the variational lower bound, both the inference neural network and MF can be learned from the data.

Efficiently learning a recommendation model from implicit feedback is also challenging since it requires to account for the large-scale unobserved data. Although stochastic gradient descent optimizer (SGD) with sampling strategy can be employed to speed up the inference, it usually suffers from low convergency and high gradient instability. Especially in the implicit recommendation task, the optimizer usually samples the uninformative data, which have low confidence and make limited contributions on gradient update \cite{chen2019samwalker}. The final recommendation performance will suffer. Instead, we turn to employ batch gradient descent optimizer (BGD), which computes stable gradient from all feedback data and usually converges to a better optimum. Unfortunately, the naive implement of BGD will suffer from low efficiency caused by the expensive full-batch gradient computation over all data. To address this problem, we develop a fast BGD-based learning algorithm fBGD for our FAWMF. With rigorous mathematical reasoning, massive repeated computations of original BGD can be avoided and the learning process can be accelerated. Notably, despite fBGD computes gradient over all data, its actual complexity is linear with the number of the unobserved positive data. Due to the sparsity of the implicit feedback data, fBGD achieves a significant acceleration.

We summarize our key contributions as follows:
\begin{itemize}
\item We propose a FAWMF model based on variational auto-encoder to achieve both adaptive weights assignment and efficient model learning.
\item A batch-based learning algorithm fBGD has been developed to learn FAWMF efficiently and stably.
\item Extensive experimental evaluations on three well-known benchmark datasets demonstrate the superiority of our FAWMF over the existing implicit MF methods.
\end{itemize}

\section{Related work}
\textbf{Recommendation from implicit feedback data.} Most of the existing methods manually assign coarse-grained confidence weights. For example, the classic weighted factorization matrix model (WMF) \cite{hu2008collaborative}, eALS \cite{he2016fast}, ICD-MF \cite{bayer2017generic}, used a simple heuristic where the confidence weights are assigned uniformly or based on item popularity; Logistical matrix factorization \cite{johnson2014logistic}, BPR \cite{rendle2009bpr}, or neural-based recommendation models (e.g. CDAE \cite{wu2016collaborative}, NCF\cite{he2017neural}) downweight the contribution of negative data implicitly in their heuristic negative sampling strategy (e.g. uniformly). More recently, a new probabilistic model EXMF\cite{liang2016modeling} was proposed to incorporate user's exposure to items into the CF methods. When inferring user's preference, user's exposure can be translated as data confidence. However, this method suffers from the efficiency problem. Also, some sophisticated models have been proposed to learn confidence weights from users' social contexts \cite{DBLP:conf/aaai/WangZYZ18,chen2018modeling,chen2019samwalker}. However, the social information is not available in many cases.

\textbf{Efficient learning algorithm for implicit recommendation models.} To handle the large-scale unobserved data, two types of strategies have been proposed for efficient learning: sample-based learning and whole-data based learning. The first type achieves fast learning with stochastic gradient descent (SGD) and negative sampling. The most popular sampling strategy is to draw un-observed feedback data uniformly, which is widely adopted in LMF\cite{johnson2014logistic}, BPR\cite{rendle2009bpr}, CDAE \cite{wu2016collaborative}, NCF \cite{he2017neural}, Mult-DAE \cite{liang2018variational} etc. Also, \cite{yu2017selection}, \cite{chen2017sampling} and \cite{hernandez2014stochastic} further propose item popularity-based and item-user co-bias sampling strategy. However, in recommendation task SGD usually suffers from slow convergence and gradient instability when the number of items is large \cite{chen2019samwalker}.

Thus, some other dynamic sampling strategies are proposed to improve convergence and accuracy. Some literatures \cite{rendle2014improving,yu2018walkranker,park2019adversarial,wang2017irgan} propose to oversample the ``difficult'' negative instances which are hard to be discriminated by the models. However, they will expose efficiency issues in sampling. Also, the stochastic gradient estimator is biased and may amplify the natural noise in users feedback data \cite{li2018adaerror}. More recently, SamWalker \cite{chen2019samwalker} conduct the random walk along the social network to select informative instances. Although effectively, it has two weakness: (1) SamWalker requires additional social information but it is not available in many recommender systems. (2) SamWalker still employs uniformly sampling strategy to update its dynamic sampler, leading to insufficient training of it.

A more effective and stable way is updating the model from the whole-data, but it face efficiency challenge due to the large number of the negative data. Thus, memorization strategies (e.g. ALS, eALS) has been proposed \cite{hu2008collaborative,he2016fast,bayer2017generic,chen2019social} to speed up learning. However, these algorithms are just suitable for the MF with simple manual confidence weights, which lacks flexible and may create empirical bias. The recommendation performance will suffer.

Adaptive weights assignment and efficient learning are both important in recommendation from implicit feedback. Existing methods are not able to provide both of these, which motivates the approach described in this paper.

\section{Preliminaries}
\label{se:exmf}
In this section, we first give the problem definition of implicit recommendation. Then, we introduce exposure-based matrix factorization (EXMF) \cite{liang2016modeling} framework from variational perspective to provide usual insight about the relation between user's exposure and data confidence.

\subsection{Problem definition}
Suppose we have a recommender system with user set $U$ (including $n$ users) and item set $I$ (including $m$ items). The implicit feedback data is represented as $n \times m$ matrix $\mathbf{X}$ with entries $x_{ij}$ denoting whether or not the user $i$ has consumed the item $j$. The task of a recommender system can be stated as follow: recommending items for each user that are most likely to be consumed by him.

\subsection{Exposure-based matrix factorization (EXMF)}
Note that unobserved feedback data $\{x_{ij}\in \mathbf{X}:x_{ij}=0\}$ contain the real negative data (dislike) and the missing values (unknown). EXMF introduces a Bernoulli variable $a_{ij}$ to model users' exposure: $a_{ij}=1$ denotes that the user $i$ knows the item $j$ and $a_{ij}=0$ denotes not. Then, EXMF models user's consumption $x_{ij}$ based on $a_{ij}$ as follow:
\begin{align}
   {a_{ij}}&\sim Bernoulli({\eta _{ij}}) \hfill \\
  ({x_{ij}}|{a_{ij}} = 1)&\sim N(\boldsymbol{u}_i^{\top}{\boldsymbol{v}_j},\lambda_x) \hfill \\
  ({x_{ij}}|{a_{ij}} = 0)&\sim {\delta _0}\approx N(\varepsilon,\lambda_x)\hfill
\end{align}
where $\delta _0$ denotes $p(x_{ij}=0|a_{ij}=0)=1$; ${\eta _{ij}}$ is the prior probability of exposure. Here we relax function $\delta _0$ as $N(\varepsilon,\lambda_x)$ to make model more robust, where $\varepsilon$ is a small constant (e.g. $\varepsilon$=1e-5). When $a_{ij}=0$, we have $x_{ij}\approx 0$, since the user does not know the item and he can not consume it. When $a_{ij}=1$, when the user has learned the item, he will decide whether or not to consume the item based on his preference. Thus, $x_{ij}$ is modeled as the classic matrix factorization model, where $\boldsymbol{u}_i$ denotes the $K$-dimensional latent factor (preference) of user $i$, and $\boldsymbol{v}_j$ denotes the $K$-dimensional latent factor (attribute) of item $j$.
\subsection{Analyses of EXMF from variational perspective}
The marginal likelihood of EXMF is composed of a sum over each datapoint $\log p(\mathbf{X}) = \sum\limits_{i\in U,j\in I} {\log p({x_{ij}})}$, which can be rewritten as:
\begin{align}
\log p({x_{ij}}) &= {E_q}[\log p({x_{ij}},{a_{ij}}) - \log q({a_{ij}}|{x_{ij}})]\notag \\
 &+ {E_q}[\log p({a_{ij}}|{x_{ij}}) - \log q({a_{ij}}|{x_{ij}})] \notag \\
 &= L(u,v,q;{x_{ij}}) + D_{KL}(q({a_{ij}}|{x_{ij}})||p({a_{ij}}|{x_{ij}}))
\end{align}
where $q(a_{ij}|x_{ij})$ is defined as an approximated variational posterior of $a_{ij}$. Since the second term -- KL-divergence -- is non-negative, the first term $L(\boldsymbol{\theta} ,q;{x_{ij}})$ is the evidence lower bound on (ELBO) the margin likelihood. Classic variational methods \cite{hoffman2013stochastic} usually employ conjugate variational distribution and individual variational parameters\footnote{Note that the EM algorithm presented in \cite{liang2016modeling} is a special case of the classic variational inference.}, i.e. $q({a_{ij}}|{x_{ij}})=Bernoulli(\gamma_{ij})$. Then, optimizing EXMF can be transferred to minimize the following objective function w.r.t $\boldsymbol{u}_i,\boldsymbol{v}_j,\gamma_{ij}$:
\begin{align}
&J(u,v ,\gamma ;\mathbf{X}) = -\frac{2}{\lambda_x} L(u,v,q;{x_{ij}}) \notag \\
& = \sum\limits_{\mathclap{i\in U,j\in I}} {{\gamma _{ij}}(\boldsymbol{u}_i^T\boldsymbol{v}_j-x_{ij})^2}  +\sum\limits_{\mathclap{i\in U,j\in I}} {({1-\gamma _{ij}})(\varepsilon-x_{ij})^2} \notag \\
&-\frac{2}{\lambda_x}\sum\limits_{{i\in U,j\in I}} {D_{KL}(q(a_{ij}|x_{ij})||p(a_{ij}))}  \notag \\
&=\sum\limits_{\mathclap{i\in U,j\in I}} {{\gamma _{ij}}(\boldsymbol{u}_i^T\boldsymbol{v}_j-x_{ij})^2}+\sum\limits_{\mathclap{i\in U,j\in I}} l(\gamma_{ij}) \label{eq:al}
\end{align}
\textbf{Exposure as data confidence.} The objective function consists of three parts: (1) Weighted matrix factorization to learn user's preference; (2) The loss when the data are predicted as unknown; (3) The regularization term -- KL divergency between the prior and the variational posterior. A good property is observed that the parameters $\gamma_{ij}$, which characterize the probability that user $i$ is exposed to item $j$, act as the confidence of the corresponding data in learning user's preference. It is consistent with our intuitions. Only if the user is exposed to the item, he can decide whether to consume the item based on his preference. The data with larger exposure are more reliable in deriving user's preference.

\textbf{Inefficiency problem.} EXMF suffers efficiency problems. The number of inferred variational parameters (confidence weights) $\gamma _{ij}$ grows quickly with the number of users and items ($n \times m$), which easily scale to billion or even larger level. It will potentially suffer from overfitting and become the time and space bottleneck for practise recommender systems. Further, the inference of EXMF requires to sum over the terms for each data, which is time-consuming.
\section{Fast adaptively weighted matrix factorization}
To address these problems, we propose a fast adaptively weighted matrix factorization (FAWMF) based on variational auto-encoder to achieve both adaptive weight assignment and efficient model learning. To reduce the number of the inferred parameters, FAWMF replaces individual parameters $\gamma _{ij}$ with a parameterized function $\gamma _{ij}=g_\Phi (i,j,\mathbf{X})$ that maps users' consumption on items into their exposure with parameters $\Phi$. It is more reasonable since there exists interactions between users. Independent assumption of exposure is not practical in real world. Typically, recent literatures \cite{palla2005uncovering} in social science claim that each of us belongs to some information-sharing communities. Users are exposed to these communities and thus their exposure exhibit commonality when they belong to similar communities. Motivated by this point, we summarize communities as collections of similar users and further infer users' exposure from the communities that they belong to. Concretely, the inference function can be modeled as follow:
\begin{align}
{g_\Phi }(i,j,\mathbf{X}) = {\boldsymbol{\theta}}_i^T\sigma ({w_j}\sum\limits_{k \in U} {{{\boldsymbol{\theta} }_k}{\alpha _k}{x_{kj}}}  + {b_j})
\end{align}

where vector $\boldsymbol{\theta}_i$ denotes the membership that allocates each user $i$ to a fixed $D$ number of communities and meet $\boldsymbol{\theta}_i \succeq 0$, $|\boldsymbol{\theta}_i|_1=1$. We cumulate the consumptions of the users in the community to infer how items are exposed to the community, where $\alpha_k$ captures the heterogenous roles of users. Different users may have different influence strength on the communities \cite{shi2019location,shi2019post}. Then, a logistic linear function with parameters $w_j,b_j$ has been employed to map cumulated consumptions into the exposure. Intuitively, the more influence users in the common community have consumed the item, the user are more likely to know it. Finally, user's exposure can be depicted by how the user belongs to the communities and how items are exposed to these communities. Overall, the inference of the data confidence can be translated into learning the parameters of inference function ${g_\Phi }(i,j,\mathbf{X})$ ($\Phi \equiv \{\boldsymbol{\theta},\alpha,w,b\}$), which captures latent correlations between $\gamma_{ij}$ for better estimation and reduces the number of inferred parameters from $O(nm)$ to $O(D(n+m))$. Different from existing clustering-based methods (e.g. \cite{chen2015wemarec,lee2016llorma}) which cluster users into communities/submatrices based on users' preference, our FAWMF deduces implicit communities based on users' exposure.

\begin{figure}[t!]
\centering
\includegraphics[width=0.46\textwidth]{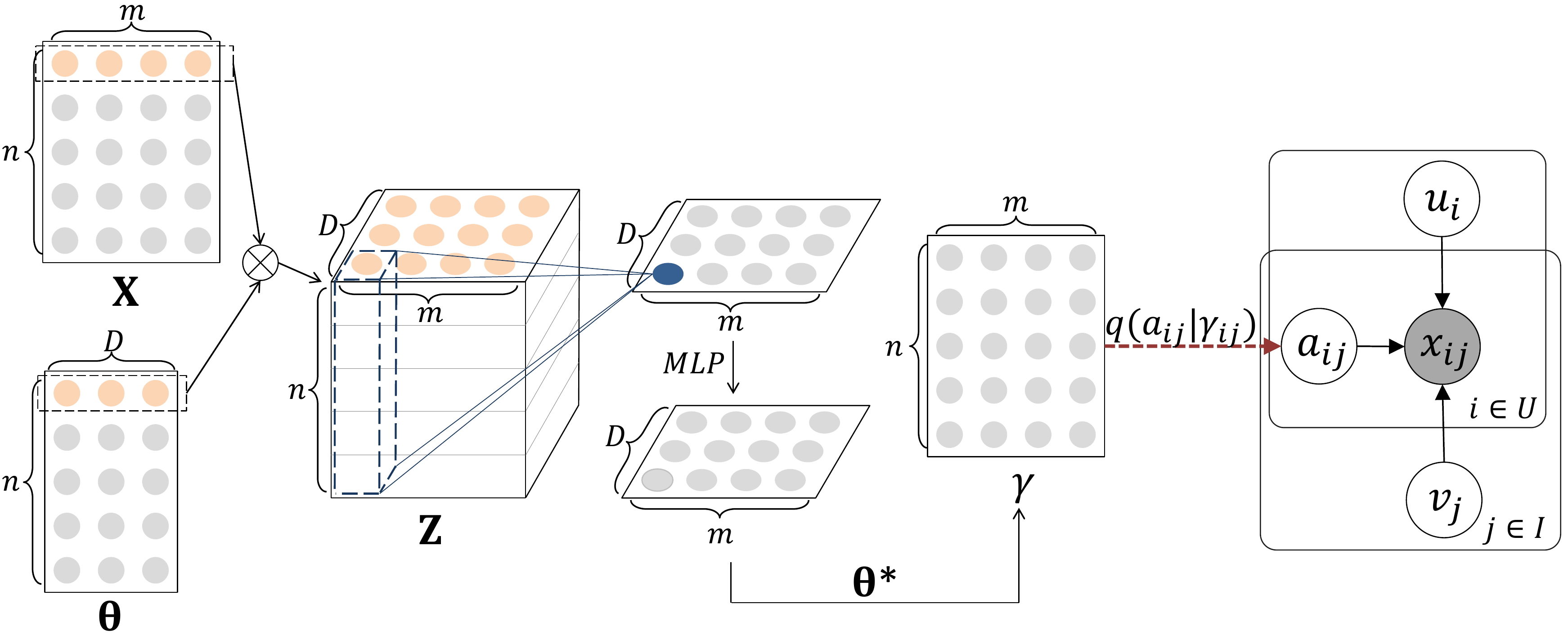}
\setlength{\abovecaptionskip}{-0.1cm}
  \setlength{\belowcaptionskip}{-0.3cm}
 \caption{Illustration of the proposed FAWMF}
\label{model}
\end{figure}

Also, FAWMF can be well understood from neural network perspective. As shown in Figure \ref{model}, at first an outer product between each user's consumption and his membership has been employed for interactions. The obtained tensor $\bf{Z}$ can be regarded as the influence of user's consumptions along different dimensions (communities). Then, we use a specific striped CNN \cite{krizhevsky2012imagenet} layer and a linear layer to encode the influences (consumptions) into the ``kernel maps'' of users' exposure. Here we choose striped CNN since the adjacent elements of $\bf{Z}$ are just caused by the adjacent users(items) id and do not suggest they will have share more commonality. Further, the exposure for different user-item pairs can be depicted as the combination of the ``kernel maps''. Finally, the exposure-based MF acts as a probabilistic decoder to predict users' consumption. Overall, FAWMF forms an auto-encoder to learn both data confidence weights and latent factors of MF.

\subsection{Discussion}
\begin{figure}[t!]
\centering
\includegraphics[width=0.46\textwidth]{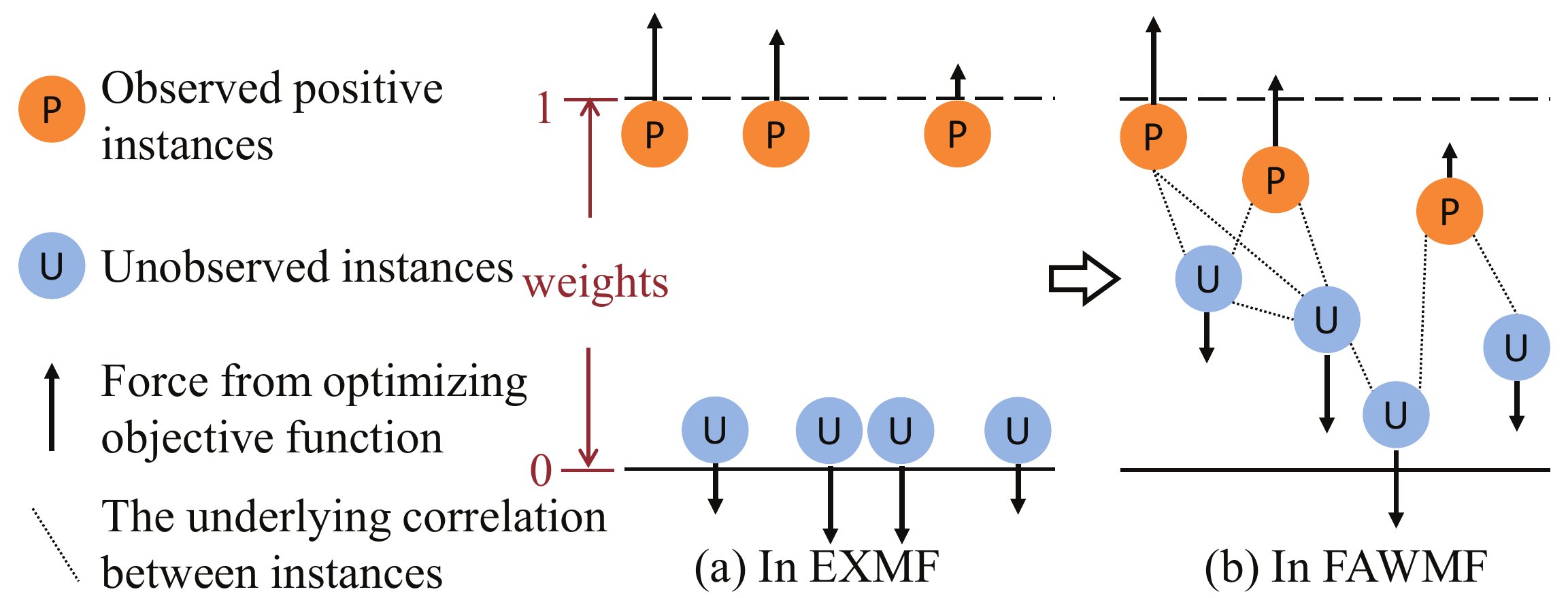}
\setlength{\abovecaptionskip}{-0.02cm}
\setlength{\belowcaptionskip}{-0.3cm}
 \caption{Illustration of how FAWMF mitigates over-fitting}
\label{shiyi}
\end{figure}

How does FAWNF mitigate overfitting? To better understand this effect, let us draw an analogy with the floating balls in the water, as illustrated in Figure \ref{shiyi}. Learning EXMF model by optimizing equation (\ref{eq:al}) will give a force to push up these positive balls (instances) and push down these unobserved balls (instances). Thus, without strong priors, the confidence weights of the data easily achieve extremely values (${\gamma _{ij}} \approx 1$ for the positive instances and ${\gamma _{ij}} \approx 0$ for the unobserved instances), where unobserved data makes few contributions on training. In this situation, all the instances will be predicted as positive by MF and the model will suffer from over-fitting. In our FAWMF, this over-fitting effect will be mitigated by introducing inference network of the ${\gamma _{ij}}$. There exists correlations between users' exposure, which can be analogies with elastic links lines between the balls. Typically, users tend to have similar exposure (${\gamma _{ij}}$), if they share more common communities. Naturally, with the training of the model, the unobserved data which have correlations with positive instances, will be pulled up due to the force from the lines. This way, when the model has well fitted the data, the positive and the unobserved instances will get stable at different depth in water. The unobserved instances which have stronger correlations with positive instances, usually reaches higher position than other unobserved instances.

\section{Fast learning algorithm fBGD}

Learning a recommendation model from implicit feedback is time-cost expensively due to the massive volume of the unobserved data. The popular solution addressing this problem is stochastic gradient descent (SGD) with negative sampling. However, SGD usually suffers from low convergency and gradient instability, which affects the performance of the model. Especially, in the recommendation task, the sampler usually select uninformative instances that have small confidence weights $\gamma_{ij}$ and make limited contributions on update. Thus, we turn to use batch gradient descent optimizer (BGD), which computes stable gradient from all feedback data. Unfortunately, the orginal BGD will
suffer from low efficiency due to the full-batch computation of the gradient over instances. To address this problem, we develop new specific learning algorithm fBGD for our FAWMF. We speed up learning of BGD by caching some specific intermediate variables to avoid the massive repeated computations. Here we detail the derivation of the gradient w.r.t $\alpha_k$; where the counterpart of others ($\boldsymbol{\theta}_i,w_j,b_j,\boldsymbol{u}_i,\boldsymbol{v}_j$) is achieved likewise and presented in supplemental materials.

For better derivation, we let $\boldsymbol{q}_j=\sigma({w_j}\sum\limits_{k \in U} {{{\boldsymbol{\theta} }_k}{\alpha _k}{x_{kj}}}  + {b_j})$. Also, we drop out the regularization term in the objective function  (Eq. (\ref{eq:al})) since the regularization term usually makes no contribution on recommendation accuracy of FAWMF and hinders fast training. It also can be regarded that we set a large value of parameter $\lambda_x$. Then, the gradient of objective function (Eq. (\ref{eq:al})) w.r.t $\alpha_k$ for each user $k$ can be derived as follow:

\begin{align}
\frac{{\partial J}}{{\partial {\alpha _k}}}& = \sum\limits_{j \in I} {\frac{{\partial J}}{{\partial {\boldsymbol{q}_j}}} \cdot \frac{{\partial {\boldsymbol{q}_j}}}{{\partial {\alpha _k}}}} \\
\frac{{\partial J}}{{\partial {\boldsymbol{q}_j}}} &= \sum\limits_{i \in U} {(\boldsymbol{u}_i^T{\boldsymbol{v}_j}\boldsymbol{u}_i^T{\boldsymbol{v}_j} - 2{x_{ij}}(\boldsymbol{u}_i^T{\boldsymbol{v}_j} - \varepsilon ) - {\varepsilon ^2}){\boldsymbol{\theta} _i}} \label{eq:dq}\\
\frac{{\partial {\boldsymbol{q}_j}}}{{\partial {\alpha _k}}} & = {\boldsymbol{q}_j} \cdot (1 - {\boldsymbol{q}_j})\cdot {\boldsymbol{\theta} _k}{w_j}{x_{kj}}
\end{align}

Clearly, the computational cost lies in two parts: (1) the calculation of the gradients of $J$ w.r.t $\boldsymbol{q}_j$ for each item $j \in I$, which requires the summation over the terms of each user $i$; (2) the summation of $\frac{{\partial J}}{{\partial {\boldsymbol{q}_j}}} \cdot \frac{{\partial {\boldsymbol{q}_j}}}{{\partial {\alpha _k}}}$ over each item $j\in I$ to get final result. In fact, the first part produces the main cost since it requires a traversal of all users and repeat this operation over all items; while the second part can be accelerated by just iterating over these positive instances ($x_{kj}=1$). The overall computational complexity to update $\{\alpha_k: k\in U\}$ is $O(nmKD$), which is generally infeasible since $nm$ can easily reach billion level or even higher in practise.

To speed up the learning, we rewrite the computational bottlenecks -- Eq. (\ref{eq:dq}) -- by isolating item-independent terms:
\begin{align}
\frac{{\partial J}}{{\partial {\boldsymbol{q}_j}}}{\rm{ }} &= \sum\limits_{k = 1}^K {\sum\limits_{l = 1}^K {{v_{jk}}{v_{jl}}\sum\limits_{i \in U} {{u_{ik}}{u_{il}}{\boldsymbol{\theta} _i}} } }  - {\varepsilon ^2}\sum\limits_{i \in U} {{\boldsymbol{\theta} _i}} \notag \\
& - 2\sum\limits_{i \in U} {{x_{ij}}(\boldsymbol{u}_i^T{\boldsymbol{v}_j} - \varepsilon ){\boldsymbol{\theta} _i}}
\end{align}
By this reformulation, we can see that the major computation -- the $\sum\limits_{i \in U} {{u_{ik}}{u_{il}}{\boldsymbol{\theta} _i}}$ and $\sum\limits_{i \in U} {{\boldsymbol{\theta} _i}}$ over all users -- is independent of item $j$. Thus, we could achieve a significant speed-up by caching these terms. That is, we cache $\mathbf{M}^{(\boldsymbol{q})}_{kl*}={\sum\limits_{i \in U} {{u_{ik}}{u_{il}}{\boldsymbol{\theta}_i}} }$ for each $1\le k\le K, 1\le l\le K$ and ${\boldsymbol{S}^{(\boldsymbol{q})}} = \sum\limits_{i \in U} {{\boldsymbol{\theta}_i}}$, where $\mathbf{M}^{(\boldsymbol{q})}$ is a $K\times K \times D$-dimensional tensor and $\boldsymbol{S}^{(\boldsymbol{q})}$ is a D-dimensional vector. Then, $\frac{{\partial J}}{{\partial {\boldsymbol{q}_j}}}$ can be calculated as follow:
\begin{align}
\frac{{\partial J}}{{\partial {\boldsymbol{q}_j}}} &= \sum\limits_{k = 1}^K {\sum\limits_{l = 1}^K {{v_{jk}}{v_{jl}}\mathbf{M}_{kl*}^{(\boldsymbol{q})}} }  - {\varepsilon ^2}{\boldsymbol{S}^{(\boldsymbol{q})}} \notag\\
&- 2\sum\limits_{i \in U} {{x_{ij}}(\boldsymbol{u}_i^T{\boldsymbol{v}_j} - \varepsilon ){\boldsymbol{\theta} _i}}  \label{eq:ndq}
\end{align}

The rearrangement of nested sums is the key transformation that allows the fast optimization. The time computation can be reduced from $O(nmKD)$ to $O(K^2D(n+m)+|\mathbf{X}^+|(K+D))$, where $|\mathbf{X}^+|$ denotes the number of the observed positive data. Here we spend $O(K^2Dn)$ time on calculating $\mathbf{M}^{(\boldsymbol{q})},\boldsymbol{S}^{(\boldsymbol{q})}$ and $O(K^2Dm)$ time on calculating the first and second term in equation (\ref{eq:ndq}) for each item $j\in I$. For the third term, we just sum over the terms with $x_{ij}=1$ with complexity $O(|\mathbf{X}^+|(K+D))$. Similar strategy can be used to accelerate the update of other parameters. With this algorithm, learning FAWMF model is efficient and stable, which trains on the all data but its complexity is linear to the number of the observed data.

\section{Experiments and analyses}
In this section, we conduct experiments to evaluate the performance of FAWMF. Our experiments are intended to address the following major questions:
\begin{enumerate}[(Q1)]
\item  Does FAWMF outperform state-of-the-art implicit MF methods?
\item  How does the proposed batch-based learning algorithm fBGD perform?
\end{enumerate}
\subsection{Experimental protocol}
\textbf{Datasets.} Three benchmark datasets Moivelens \footnote{\url{https://grouplens.org/datasets/movielens/}}, Amazon (food-reviews) \footnote{\url{https://www.kaggle.com/snap/amazon-fine-food-reviews}}, Douban \footnote{\url{https://www.cse.cuhk.edu.hk/irwin.king.new/pub/data/douban}} are used in our experiments. These datasets contain users' feedback on items. The dataset statistics are presented in Table \ref{tb:da}. Similar to \cite{chen2019samwalker}, we preprocess the datasets so that all items have at least three interactions and``binarize'' user's feedback into implicit feedback. That is, as long as there exists some user-item interactions (ratings or reviews), the corresponding implicit feedback is assigned a value of 1. Grid search and 5-fold cross validation are used to find the best parameters. In our FAWMF, we set $D$=$K$=$20$, $\varepsilon$=1e-5 and learning rate=$0.1$ across all datasets.

\begin{table}[t!]
\centering
\scriptsize
\caption{Statistics of three datasets.}
\label{tb:da}
\begin{tabular}{|c|c|c|c|}
\hline
Datasets & \#Users & \#Items &   \#Oberseved positive feedback  \\ \hline
Movielens   & 6,040   & 3,952   & 1,000,209                    \\ \hline
Amazon     & 10,619   &  37,762  &  256,287                  \\ \hline
Douban & 123,480  &   20,029  & 16,624,937                  \\ \hline

\end{tabular}
\end{table}

\begin{table}[t!]
\scriptsize
\caption{characteristics of the compared methods.}
\label{tb:ch}
\begin{tabular}{|c|c|c|c|}
\hline
Methods  & \begin{tabular}[c]{@{}c@{}}Adaptive\\ weights?\end{tabular} & \begin{tabular}[c]{@{}c@{}}Without\\ sampling?\end{tabular} & Complexity                   \\ \hline
WMF(ALS) & $\backslash$                                                & ${\surd}$                                                   & $\scriptstyle O((n+m)K^3+|\mathbf{X}^+|K^2)$                \\ \hline
eALS     & $\backslash$                                                & ${\surd}$                                                   & $\scriptstyle O((n+m)K^2+|\mathbf{X}^+|K)$         \\ \hline
BPR      & $\backslash$                                                & $\backslash$                                                & $\scriptstyle O((n+m+|\mathbf{X}^+|)K)$            \\ \hline
CDAE     & $\backslash$                                                & $\backslash$                                                & $\scriptstyle O((n+m+|\mathbf{X}^+|)K)$            \\ \hline
EXMF     & ${\surd}$                                                   & ${\surd}$                                                   & $\scriptstyle O(nmK^3)$                     \\ \hline
FAWMF    & ${\surd}$                                                   & ${\surd}$                                                   & $\scriptstyle O((n+m)K^2D+|\mathbf{X}^+|(K+D))$ \\ \hline
\end{tabular}
\end{table}

\begin{table*}[t!]
\scriptsize
\center
\caption{The performance metrics of the compared methods. The boldface font denotes the winner in that column.  The row `Impv' indicates the relative performance gain of our FAWMF compared to the best results among baselines. '$\dagger$' indicates that the improvement is significant with t-test at $p < 0.05$.}
\label{tb:ans}
\begin{tabular}{|c|c|c|c|c|c|c|c|c|c|c|c|c|}
\hline
\multirow{2}{*}{Methods} & \multicolumn{4}{c|}{Movielens}                                        & \multicolumn{4}{c|}{Amazon}                                              & \multicolumn{4}{c|}{Douban}                                             \\ \cline{2-13}
                         & Pre@5           & Rec@5           & NDCG@5          & MRR             & Pre@5            & Rec@5           & NDCG@5           & MRR              & Pre@5            & Rec@5           & NDCG@5           & MRR             \\ \hline
Item-pop                 & 0.2092          & 0.0400          & 0.2201          & 0.8958          & 0.0027           & 0.0048          & 0.0055           & 0.0191           & 0.1409           & 0.0332          & 0.1582           & 0.6308          \\ \hline
WMF(ALS)                 & 0.3841          & 0.0924          & 0.4059          & 1.5751          & 0.0789           & 0.0406          & 0.0858           & 0.3269           & 0.2400           & 0.0656          & 0.2598           & 1.0113          \\ \hline
eALS                     & 0.3955          & 0.0917          & 0.4175          & 1.5998          & 0.0984           & 0.0348          & 0.1051           & 0.3951           & 0.2329           & 0.0646          & 0.2520           & 0.9880          \\ \hline
BPR                      & 0.3613          & 0.0798          & 0.3794          & 1.5023          & 0.0988           & 0.0469          & 0.1060           & 0.3969           & 0.2371           & 0.0582          & 0.2570           & 1.0093          \\ \hline
CDAE                     & 0.3786          & 0.0860          & 0.3950          & 1.5454          & 0.0948           & \textbf{0.0472} & 0.0947           & 0.3994           & 0.2377           & 0.0589          & 0.2573           & 1.0162          \\ \hline
EXMF                     & 0.3871          & 0.0936          & 0.4071          & 1.5720          & 0.0847           & 0.0418          & 0.0928           & 0.3683           & 0.2353           & 0.0666          & 0.2588           & 1.0016          \\ \hline
FAWMF                    & \textbf{0.4054} & \textbf{0.0949} & \textbf{0.4275} & \textbf{1.6279} & \textbf{0.1129}  & 0.0441          & \textbf{0.1285}  & \textbf{0.4470}  & \textbf{0.2661}  & \textbf{0.0680} & \textbf{0.2915}  & \textbf{1.0984} \\ \hline
Impv\%                   & 2.49\%$\dagger$ & 1.41\%$\dagger$ & 2.40\%$\dagger$ & 1.76\%$\dagger$ & 14.34\%$\dagger$ & -6.42\%         & 21.18\%$\dagger$ & 11.93\%$\dagger$ & 10.88\%$\dagger$ & 2.11\%$\dagger$ & 12.19\%$\dagger$ & 8.09\%$\dagger$ \\ \hline
\end{tabular}
\end{table*}

\textbf{Compared methods.} We compare FAWMF with following baselines. Table \ref{tb:ch} concludes their characteristics.
\begin{itemize}
\item Item-pop: This is a simple baseline which recommends items based on global item popularity.
\item WMF(ALS)  \cite{hu2008collaborative,Pan2008}: The classic weighted matrix factorization model for implicit feedback data. The corresponding ALS-based  \cite{hu2008collaborative} algorithm can reduce inference complexity.
\item eALS  \cite{he2016fast}: The improved weighted matrix factorization model, where the data confidence weights are assigned heuristically based on item-popularity.
\item BPR \cite{rendle2009bpr}: The classic pair-wise method for recommendation, coupled with matrix factorization. BPR implicitly downweights the unobserved data in their uniformly negative strategy.
\item CDAE \cite{wu2016collaborative}: The advanced recommendation method based on Auto-Encoders, which is a generalization of WMF with more flexible components. CDAE also employs uniformly negative strategy to learn their model.
\item EXMF \cite{liang2016modeling}: A probabilistic model that directly incorporates user's exposure to items into traditional matrix factorization.
\end{itemize}

\textbf{Evaluation Metrics.} We adopt four well-known metrics Precision@K (Pre@K), Recall@K (Rec@K), Normalized Discounted Cumulative Gain (NDCG@K) and Mean Reciprocal Rank (MRR) to evaluate recommendation performance:  Recall@K (Rec@K) quantifies the fraction of consumed items that are in the top-K ranking list; Precision@K (Pre@K) measures the fraction of the top-K items that are indeed consumed by the user; NDCG@K and MRR evaluate ranking performance of the methods. Refer to our supplemental material for more details about these metrics.

\subsection{Performance comparison (Q1)}
Table \ref{tb:ans} presents the performance of the compared methods in terms of three evaluation metrics. The boldface font denotes the winner in that column. For the sake of clarity, the last row of Table \ref{tb:ans} also show the relative improvements achieved by our FAWMF over the baselines. Generally speaking, with one exception, FAWMF outperforms all compared methods on all datasets for all metrics.

The improvement of FAWMF over these baselines can be attributed to three aspects: (1) In the real world, users have personalized communities and thus are exposed to diverse information. Correspondingly, the data confidence (exposure) will vary with different user-item combinations. That is, some unobserved feedback are more likely attributed to user's preference while others are the results of users' awareness. By adaptively learning fine-grained data confidence weights from the data, FAWMF achieves better performance than those baselines with manually coarse-grained confidence weights. (2) FAWMF models confidence weight with an community-based inference network, which is capable of capturing latent interactions between users and mitigating the over-fitting problems. It can bee seen from the better performance of FAWMF over EXMF. (3) Instead of employing sampling-based stochastic gradient descent optimizer (SGD), FAWMF employs a specific fast batch-based learning algorithm, which has better convergency and more stable results. We also conduct specific experiments in the next subsection to validate this point.

\begin{figure}[t]
  \centering
  \subfigure[On movielens]{
\includegraphics[width=0.14\textwidth]{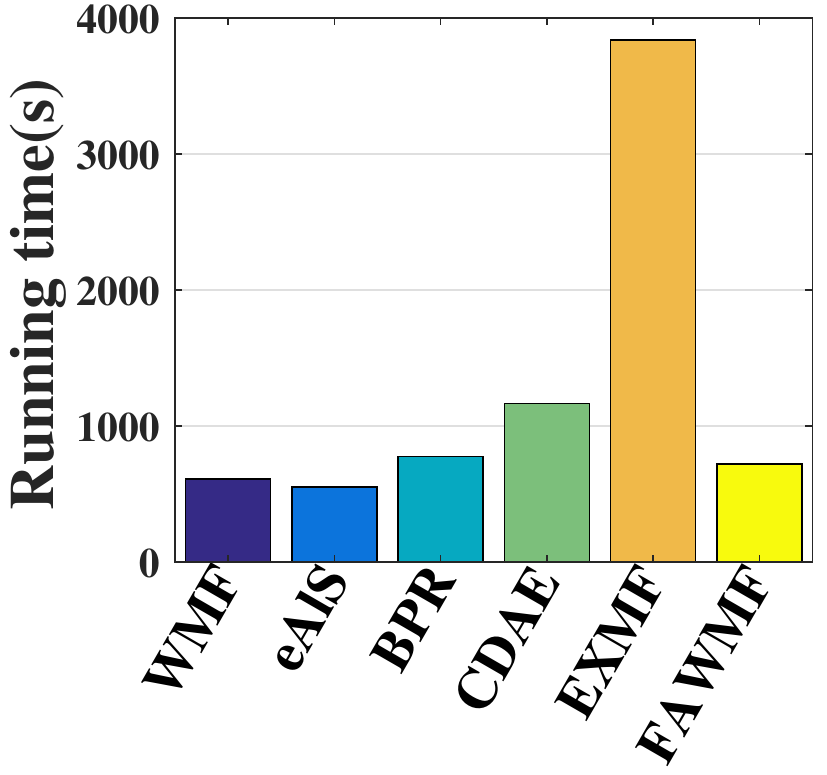}
}
\subfigure[On Amazon]{
\includegraphics[width=0.14\textwidth]{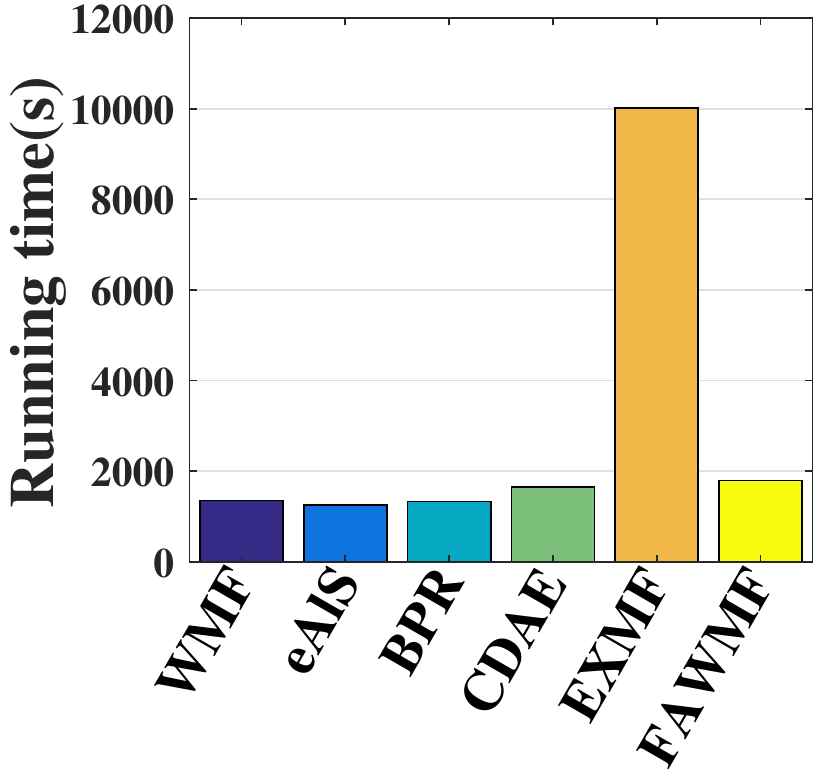}
}
\subfigure[On Douban]{
\includegraphics[width=0.14\textwidth]{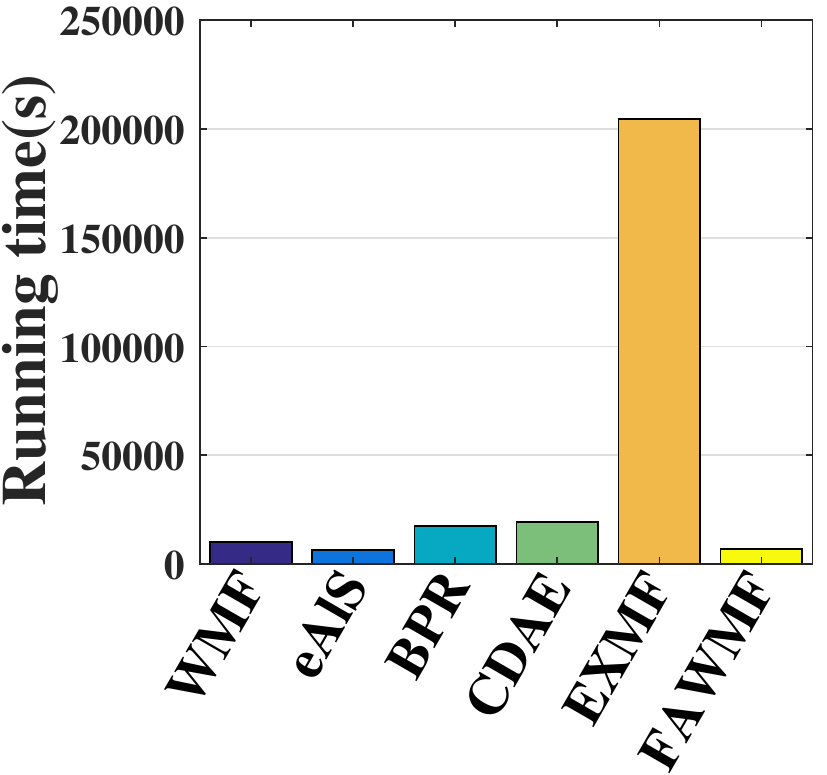}
}
\setlength{\abovecaptionskip}{-0.1cm}
 \setlength{\belowcaptionskip}{-0.2cm}
  \caption{Running time comparisons.}\label{fg:rt}
\end{figure}

\textbf{Running time comparisons}. Figure \ref{fg:rt} depicts running time of the six compared recommendation methods. As we can see, the speed up of our FAWMF over EXMF is significant. Especially in the largest dataset Douban, EXMF requires 56 hours for training, while FAWMF only takes 1.8 hours. The acceleration of FAWMF over EXMF can be attributed two aspects: (1) Confidence weights for each data have been modeled with a parameterized function, which reduces the number of learned parameters. (2) fBGD speed up gradients calculations. This way, FAMWF achieves similar analytical time complexity as other compared methods, which aim at fast learning but sacrifice the flexibility of the confidence weights. Their actual running time are also in the same magnitude. Also, we observe that these BGD-based methods (WMF, eALS, FAWMF) are relatively more efficient than SGD-based methods (BPR, CDAE), although they have similar analytical time complexity. It is caused by the poor convergency of SGD, which usually requires more iterations.

\subsection{Effect of batch-based learning algorithm (Q2)}

In this subsection, we compare our fBGD with two popular SGD-based learning strategies: (1) Uniform-1X (Uniform-5X, Uniform-25X), which uniformly sample a part of unobserved instances to update the model. Note that the training time and prediction accuracy are largely determined by the size of negative samples. Here we test this learning strategy with three different sampling-size for comparisons, where Uniform-25X (Uniform-5X, Uniform-1X) denotes the number of the sampled instances is 25 (5, 1) times as large as the number of positive instances. (2) Itempop-1X (Itempop-5X, Itempop-25X) whose probability of sampling a un-observed data is proportional to the item popularity. We also present the performance of original BGD algorithm for running time comparisons. Here we do not compare with existing dynamic sampling strategies, since they either suffer from efficiency problems or require other side information.

\begin{figure}[t!]
\centering
\includegraphics[width=0.45\textwidth]{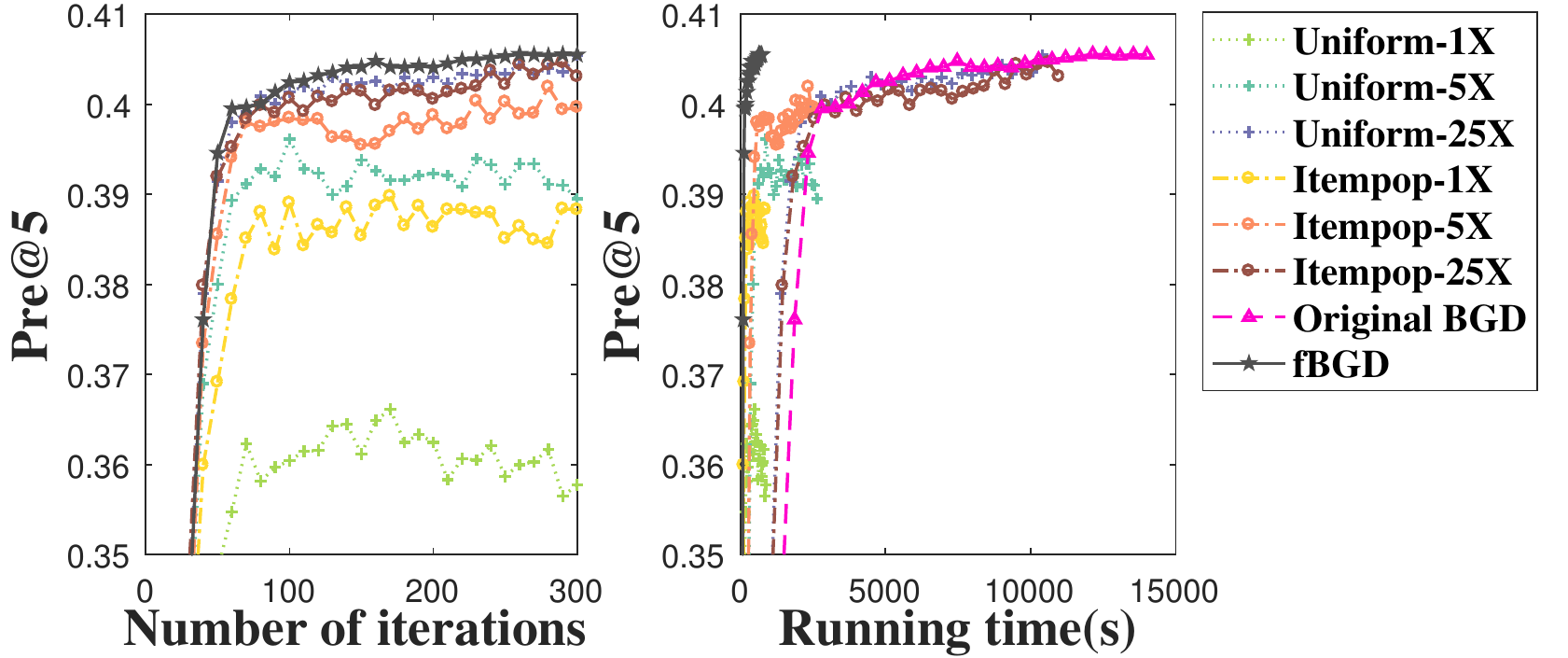}
\setlength{\abovecaptionskip}{-0.1cm}
 \setlength{\belowcaptionskip}{-0.4cm}
 \caption{Recommendation accuracy for different learning strategy versus the number of iterations and running time.}

\label{prevst}
\end{figure}

Figure \ref{prevst} presents pre@5 of FAWMF on dataset Movielens with different learning strategies versus the number of iteration and running time. There exists the efficiency and effectiveness trade-off of SGD-based learning algorithm with varying sampling-size. The smaller sampling-size will cause insufficient learning and gradient instability. It can be seen from the poor performance final and heavy fluctuation of Uniform-1X (Itempop-1X). On the contrary, when the sampling-size become larger, SGD performs better but spend much more time. However, even if the stochastic optimizer uses a large sampling-size, it can not achieve good performance as BGD-based learning algorithm. Also, we can observe the inefficiency of original BGD. Our proposed fBGD accelerates original BGD for 16 times and even performs more efficient than Itempop-1X. Overall, our proposed fBGD performs better than others in all convergence, speed and recommended performance.

\subsection{Effect of the parameter $K$}

\begin{figure}[t]
  \centering
  \subfigure[On movielens]{
\includegraphics[width=0.14\textwidth]{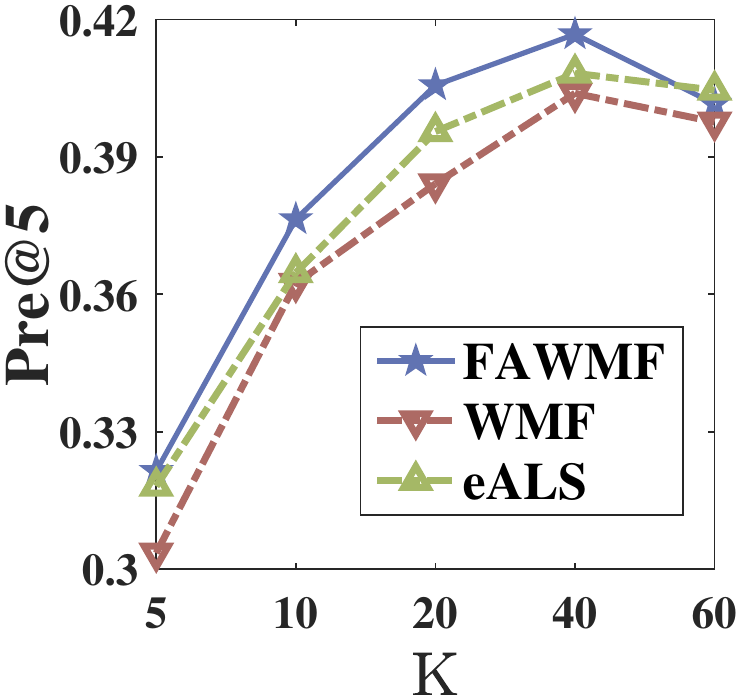}
}
\subfigure[On Amazon]{
\includegraphics[width=0.14\textwidth]{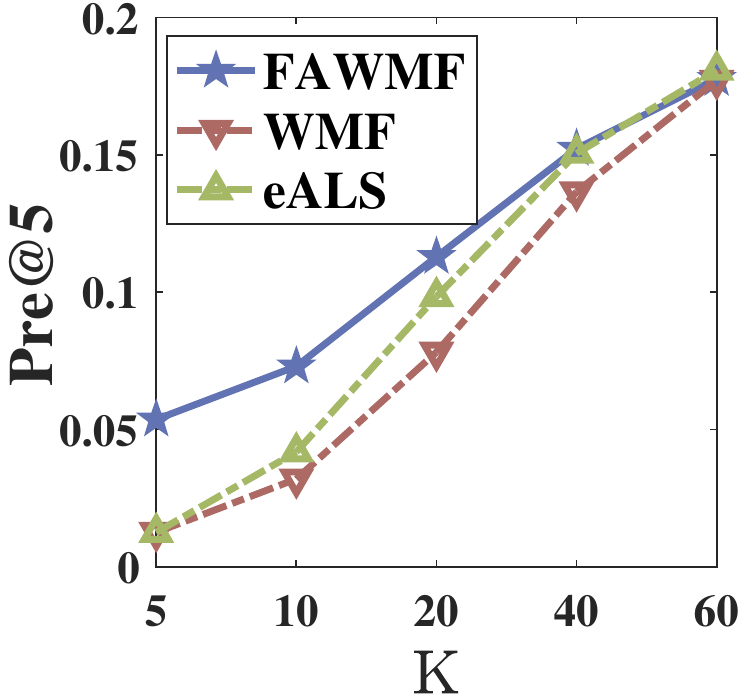}
}
\subfigure[On Douban]{
\includegraphics[width=0.14\textwidth]{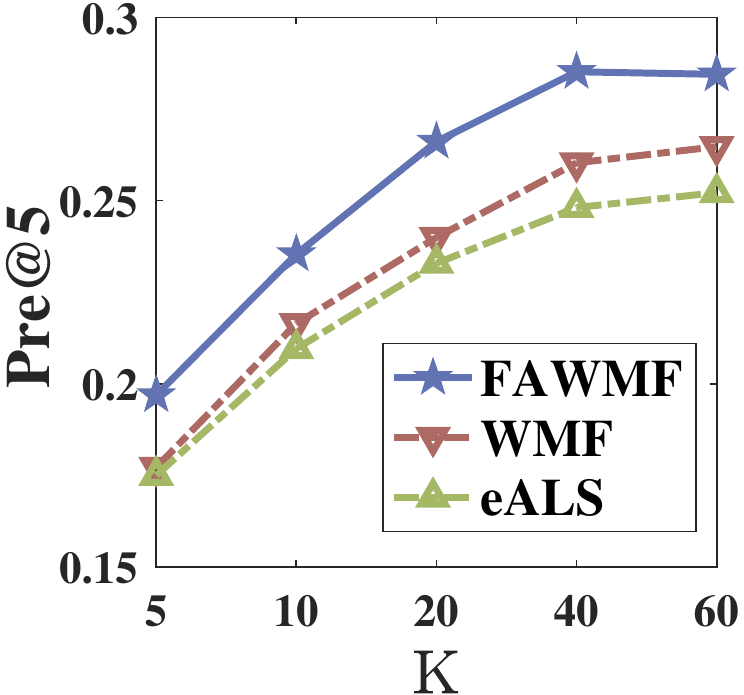}
}
\setlength{\abovecaptionskip}{-0.1cm}
 \setlength{\belowcaptionskip}{-0.2cm}
  \caption{Impact of the parameter $K$.}\label{fg:k}
\end{figure}

Figure \ref{fg:k} shows the performance of FAWMF with varying latent dimension $K$ in MF. We also present the results of two close MF methods for comparisons. First, with few exceptions, FAWMF consistently outperforms eALS and WMF across $K$, demonstrating the effectiveness of adaptively assigning personalized confidence weights. Second, all methods can be improved with a larger $K$. But a large $K$ might have the risk of overfitting. It can be seen from the worse results when $K=60$ on the dataset Movielens.

\subsection{Case study}

\begin{figure}[t!]
\centering
\includegraphics[width=0.47\textwidth]{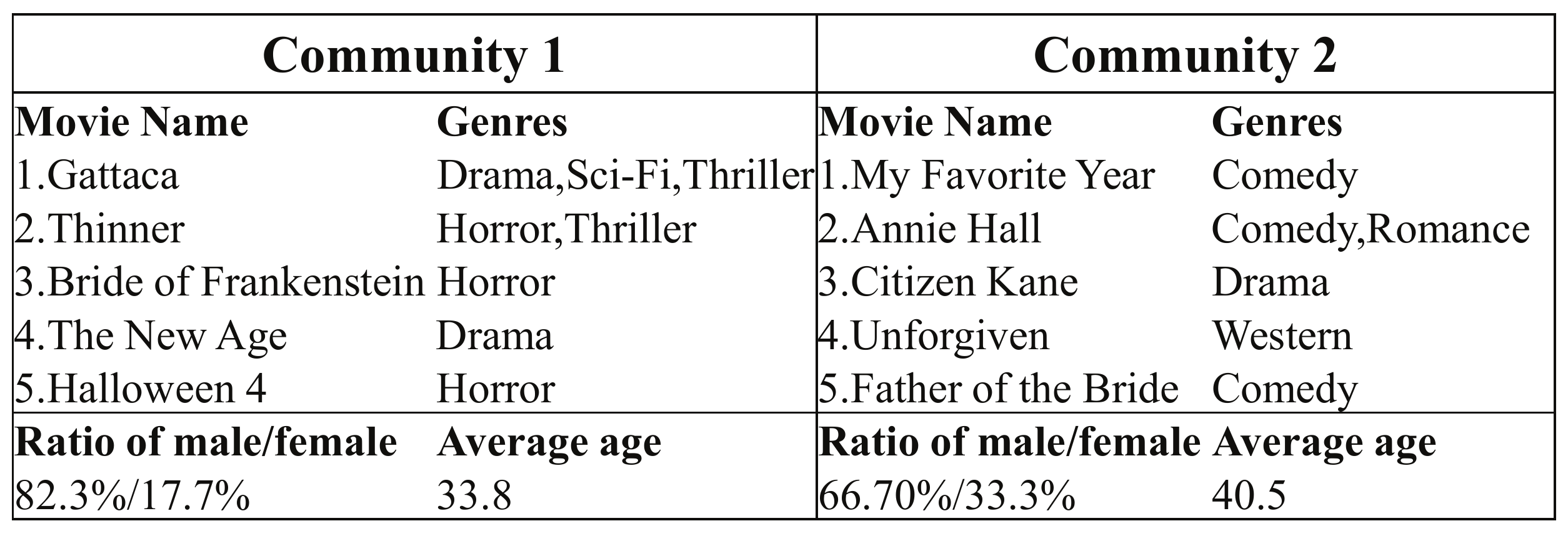}
\setlength{\abovecaptionskip}{-0.1cm}
 \setlength{\belowcaptionskip}{-0.2cm}
 \caption{Case study of two communities: the top rows show top-5 items that have highest exposure to the users in the two communities; the bottom rows show the ratio of male/female and the average age of the users in each communities. Note that these side information is not used in model training.}
\label{case}
\end{figure}

We also conduct a case study to show the learned communities. We do statistical analyses on the communities, two of which are presented in Figure \ref{case}. We can find that the genres of the exposed items in the community 1 are mainly about ``Horror'' and ``Thriller'', but the genres in the community 2 are mainly about ``Comedy''. These results can be explained by the different kinds of users in the communities. Most of users in the community 1 are young men, who are more likely to enjoy these exciting movies and share these movies with each other. Meanwhile, comedy is a hot topic for the elderly people, who are the main constituents of the community 2. These results validate the effectiveness of our FAWMF. Although the genre information of users and items is not used in model training, the latent correlations between users/items can be captured. Also, we can find different users will belong to different communities and thus have different exposure. Personalized confidence weights are necessary for the implicit recommendation models.
\section{Conclusion}
In this paper, we present a novel recommendation method FAWMF based on variational auto-encoder to achieve both adaptive weights assignment and efficient model learning. On the one hand, FAWMF models data confidence weights with a community-based inference neural network, which reduces the number of inferred parameters and is capable of capturing latent interactions between users. On the other hand, a specific batch-based learning algorithm fBGD has been developed to learn FAWMF fast and stably. The experimental results on three real-world datasets demonstrate the superiority of FAWMF over existing implicit MF methods.
\section*{ACKNOWLEDGMENTS}
This work is supported by National Natural Science Foundation of China (Grant No: U1866602) and National Key Research and Development Project (Grant No: 2018AAA0\\101503).
\bibliographystyle{aaai}
\bibliography{sigproc}

\newpage
\section{Appendix}
\subsection{Details of fBGD}
\allowdisplaybreaks[4]
Here we present the detailed derivation of our fBGD. Note that the constraint of the $\boldsymbol{\theta}$ ($\boldsymbol{\theta}_i \succeq 0$, $|\boldsymbol{\theta}_i|_1=1$)
 causes the optimization problem. We re-parameterize $\boldsymbol{\theta}_i$ by a softmax function with new parameters $\boldsymbol{\beta}_i$: $\boldsymbol{\theta}_i=s(\boldsymbol{\beta}_i)$.  The gradient of objective function w.r.t the parameters $\boldsymbol{\beta}_i$, $w_j$, $b_j$, $\alpha_k$, $\boldsymbol{u}_i$, $\boldsymbol{v}_j$ can be derived as follows:
\begin{align}
\frac{{\partial J}}{{\partial {\boldsymbol{\beta} _i}}} &= \frac{{\partial {\boldsymbol{\theta} _i}}}{{\partial {\boldsymbol{\beta} _i}}}\frac{{\partial J}}{{\partial {\boldsymbol{\theta} _i}}}\\
\frac{{\partial J}}{{\partial {w_j}}} &= \frac{{\partial J}}{{\partial {\boldsymbol{q}_j}}} \cdot \frac{{\partial {\boldsymbol{q}_j}}}{{\partial {w_j}}}\\
\frac{{\partial J}}{{\partial {b_j}}} &= \frac{{\partial J}}{{\partial {\boldsymbol{q}_j}}} \cdot \frac{{\partial {\boldsymbol{q}_j}}}{{\partial {b_j}}}\\
\frac{{\partial J}}{{\partial {\alpha _k}}} &= \sum\limits_{j \in I} {\frac{{\partial J}}{{\partial {\boldsymbol{q}_j}}} \cdot \frac{{\partial {\boldsymbol{q}_j}}}{{\partial {\alpha _k}}}} \\
\frac{{\partial J}}{{\partial {\boldsymbol{u}_i}}} &= 2\sum\limits_{j \in I} {\boldsymbol{\theta} _i^T{\boldsymbol{q}_j}({\boldsymbol{v}_j}\boldsymbol{u}_i^T{\boldsymbol{v}_j} - {x_{ij}}{\boldsymbol{v}_j})} \\
\frac{{\partial J}}{{\partial {\boldsymbol{v}_j}}} &= 2\sum\limits_{i \in U} {\boldsymbol{\theta} _i^T{\boldsymbol{q}_j}({u_j}\boldsymbol{u}_i^T{\boldsymbol{v}_j} - {x_{ij}}{\boldsymbol{u}_i})} \\
\frac{{\partial J}}{{\partial {\boldsymbol{\theta} _i}}} &= \sum\limits_{j \in I}{(\boldsymbol{u}_i^T{\boldsymbol{v}_j}\boldsymbol{u}_i^T{\boldsymbol{v}_j} - 2{x_{ij}}({\mathop{\rm u}\nolimits} _i^T{\boldsymbol{v}_j} - \varepsilon ) - {\varepsilon ^2}){\boldsymbol{q}_j}} \notag \\
 &+ \sum\limits_{j \in I} {\frac{{\partial {\boldsymbol{q}_j}}}{{\partial {\boldsymbol{\theta} _i}}}\frac{{\partial J}}{{\partial {\boldsymbol{q}_j}}}} \\
\frac{{\partial J}}{{\partial {\boldsymbol{q}_j}}} &= \sum\limits_{i \in U} {(\boldsymbol{u}_i^T{\boldsymbol{v}_j}\boldsymbol{u}_i^T{\boldsymbol{v}_j} - 2{x_{ij}}(\boldsymbol{u}_i^T{\boldsymbol{v}_j} - \varepsilon ) - {\varepsilon ^2}){\boldsymbol{\theta} _i}} \\
\frac{{\partial {\boldsymbol{q}_j}}}{{\partial {\boldsymbol{\theta} _i}}}& = {\bf{I}}({\boldsymbol{q}_j} \cdot (1 - {\boldsymbol{q}_j}){\alpha _i}{w_j}{x_{ij}})\\
\frac{{\partial {\boldsymbol{\theta} _i}}}{{\partial {\boldsymbol{\beta} _i}}} &= {\bf{I}}{\boldsymbol{\theta} _i} - {\boldsymbol{\theta} _i}\boldsymbol{\theta} _i^T\\
\frac{{\partial {\boldsymbol{q}_j}}}{{\partial {w_j}}} &= {\boldsymbol{q}_j} \cdot (1 - {\boldsymbol{q}_j}) \cdot \sum\limits_{k \in U} {{\boldsymbol{\theta} _k}{\alpha _k}{x_{kj}}} \\
\frac{{\partial {\boldsymbol{q}_j}}}{{\partial {b_j}}} &= {\boldsymbol{q}_j} \cdot (1 - {\boldsymbol{q}_j})\\
\frac{{\partial {\boldsymbol{q}_j}}}{{\partial {\alpha _k}}} &= {\boldsymbol{q}_j} \cdot (1 - {\boldsymbol{q}_j}) \cdot {\boldsymbol{\theta} _k}{w_j}{x_{kj}}
\end{align}

As we can see, the computational bottlenecks is on computing $\frac{{\partial J}}{{\partial {\boldsymbol{\theta} _i}}}$ (Eq.(7)), $\frac{{\partial J}}{{\partial {\boldsymbol{q}_j}}}$ (Eq.(8)), $\frac{{\partial J}}{{\partial {\boldsymbol{u}_i}}}$ (Eq.(5)), $\frac{{\partial J}}{{\partial {\boldsymbol{v}_j}}}$ (Eq.(6)) since it requires a traversal of all users (items) and repeat this operation over all items (users); while others can be accelerated by just iterating over these positive instances ($x_{ij}=1$). The overall computational complexity to update $\alpha$ is $O(nmKD$), which is generally infeasible since $nm$ can easily reach billion level or even higher in practise.

To speed up the learning, we first rewrite the computational bottlenecks --Eq.(5)(6)(7)(8) by isolating item(user)-independent terms as follows:

\begin{align}
\frac{{\partial J}}{{\partial {\boldsymbol{\theta} _i}}} &= \sum\limits_{k = 1}^D {\sum\limits_{l = 1}^D {{u_{ik}}{u_{il}}\sum\limits_{j \in I} {{v_{jk}}{v_{jl}}{\boldsymbol{q}_j}} } }  - {\varepsilon ^2}\sum\limits_{j \in I} {{\boldsymbol{q}_j}} \notag \\
& - 2\sum\limits_{j \in I} {{x_{ij}}(\boldsymbol{u}_i^T{\boldsymbol{v}_j} - \varepsilon ){\boldsymbol{q}_j}}  + \sum\limits_{j \in I} {\frac{{\partial {\boldsymbol{q}_j}}}{{\partial {\boldsymbol{\theta} _i}}}\frac{{\partial J}}{{\partial {\boldsymbol{q}_j}}}} \\
\frac{{\partial J}}{{\partial {\boldsymbol{q}_j}}} &= \sum\limits_{k = 1}^D {\sum\limits_{l = 1}^D {{v_{jk}}{v_{jl}}\sum\limits_{i \in U} {{u_{ik}}{u_{il}}{\boldsymbol{\theta} _i}} } }  - {\varepsilon ^2}\sum\limits_{i \in U} {{\boldsymbol{\theta} _i}} \notag \\
& - 2\sum\limits_{i \in U} {{x_{ij}}(\boldsymbol{u}_i^T{\boldsymbol{v}_j} - \varepsilon ){\boldsymbol{\theta} _i}} \\
\frac{{\partial J}}{{\partial {\boldsymbol{u}_i}}} &= 2\sum\limits_{k = 1}^D {\sum\limits_{l = 1}^K {{\theta_{ik}}{u_{il}}\sum\limits_{j \in I} {{q_{jk}}{v_{jl}}{\boldsymbol{v}_j}} } }  - 2\sum\limits_{j \in I} {\boldsymbol{\theta} _i^T{\boldsymbol{q}_j}{x_{ij}}{\boldsymbol{v}_j}} \\
\frac{{\partial J}}{{\partial {\boldsymbol{v}_j}}} &= 2\sum\limits_{k = 1}^D {\sum\limits_{l = 1}^K {{q_{jk}}{v_{jl}}\sum\limits_{i \in U} {{\theta_{ik}}{u_{il}}{\boldsymbol{u}_i}} } }  - 2\sum\limits_{i \in U} {\boldsymbol{\theta} _i^T{\boldsymbol{q}_j}{x_{ij}}{\boldsymbol{u}_i}}
\end{align}

By this reformulation, we can see that the major computation -- the $\sum\limits_{i \in U} {{u_{ik}}{u_{il}}{\boldsymbol{\theta} _i}}$, $\sum\limits_{i \in U} {\boldsymbol{\theta} _i}$, $\sum\limits_{i \in U} {{\boldsymbol{\theta}_{ik}}{u_{il}}{\boldsymbol{u}_i}} $ over all users -- is independent of item $j$; and the$\sum\limits_{j \in I} {{v_{jk}}{v_{jl}}{\boldsymbol{q}_j}}$, $\sum\limits_{j \in I} {{\boldsymbol{q}_j}}$, $\sum\limits_{j \in I} {{q_{jk}}{v_{jl}}{\boldsymbol{v}_j}} $ over all items is independent of user $i$. Thus, we could achieve a significant speed-up by caching these terms. That is, we cache:
\begin{align}
M^{(\boldsymbol{\theta})}_{kl*}&={\sum\limits_{j \in I} {{v_{jk}}{v_{jl}}{\boldsymbol{q}_j}} } \\
{S^{(\boldsymbol{\theta})}}& = \sum\limits_{j \in I} {{\boldsymbol{q}_j}} \\
M_{kl*}^{(q)} &= \sum\limits_{i \in U} {{u_{ik}}{u_{il}}{\boldsymbol{\theta}_i}} \\
{S^{(q)}} &= \sum\limits_{i \in U} {{\boldsymbol{\theta} _i}}
\end{align}
for each $1\le k\le D, 1\le l\le D$ and:
\begin{align}
M_{kl*}^{(u)}& = \sum\limits_{j \in I} {{q_{jk}}{v_{jl}}{\boldsymbol{v}_j}} \\
M_{kl*}^{(v)} &= \sum\limits_{i \in U} {{\theta_{ik}}{u_{il}}{\boldsymbol{u}_i}}
\end{align}
for each $1\le k\le D, 1\le l\le K$. where $M^{(\boldsymbol{\theta})}$, $M^{(q)}$ are $K\times K \times D$-dimensional tensors and $S^{(\boldsymbol{\theta})}$, $S^{(q)}$ are $D$-dimensional vectors; $M^{(u)}$, $M^{(v)}$ are $D\times K \times K$-dimensional tensors and $S^{(\boldsymbol{\theta})}$, $S^{(q)}$ are $K$-dimensional vectors. This way, the computational bottlenecks Eq.(5)(6)(7)(8) can be efficiently calculated as follows:

\begin{align}
\frac{{\partial J}}{{\partial {\boldsymbol{\theta} _i}}} &= \sum\limits_{k = 1}^D {\sum\limits_{l = 1}^D {{u_{ik}}{u_{il}}M_{kl*}^{(\boldsymbol{\theta} )}} }  - {\varepsilon ^2}{S^{(\boldsymbol{\theta} )}}\notag \\
 &- 2\sum\limits_{j \in I} {{x_{ij}}(\boldsymbol{u}_i^T{\boldsymbol{v}_j} - \varepsilon ){\boldsymbol{q}_j}}  + \sum\limits_{j \in I} {\frac{{\partial {\boldsymbol{q}_j}}}{{\partial {\boldsymbol{\theta} _i}}}\frac{{\partial J}}{{\partial {\boldsymbol{q}_j}}}} \\
\frac{{\partial J}}{{\partial {\boldsymbol{q}_j}}} &= \sum\limits_{k = 1}^D {\sum\limits_{l = 1}^D {{v_{jk}}{v_{jl}}M_{kl*}^{(q)}} }  - {\varepsilon ^2}{S^{(q)}}\notag \\
& - 2\sum\limits_{i \in U} {{x_{ij}}(\boldsymbol{u}_i^T{\boldsymbol{v}_j} - \varepsilon ){\boldsymbol{\theta} _i}} \\
\frac{{\partial J}}{{\partial {\boldsymbol{u}_i}}} &= 2\sum\limits_{k = 1}^D {\sum\limits_{l = 1}^K {{\theta_{ik}}{u_{il}}M_{kl*}^{(u)}} }  - 2\sum\limits_{j \in I} {\boldsymbol{\theta} _i^T{\boldsymbol{q}_j}{x_{ij}}{\boldsymbol{v}_j}} \\
\frac{{\partial J}}{{\partial {\boldsymbol{v}_j}}} &= 2\sum\limits_{k = 1}^D {\sum\limits_{l = 1}^K {{q_{jk}}{v_{jl}}M_{kl*}^{(v)}} }  - 2\sum\limits_{i \in U} {\boldsymbol{\theta} _i^T{\boldsymbol{q}_j}{x_{ij}}{\boldsymbol{u}_i}}
\end{align}

The rearrangement of nested sums and the cache strategies can avoid the massive repeated computations. The time computation can be reduced from $O(nmKD)$ to $O((n+m)K^2D+|X^+|(K+D))$. That is, although our fBGD uses all feedback data but its complexity is linearly to the number of observed data. due to the sparsity of the implicit feedback data, our FAWMF with fBGD learning algorithm is efficient. Our experimental results also validate this point. Overall, our fBGD algorithm is presented in algorithm 1.

\begin{algorithm}[t!]
\small
\caption{Inference of FAWMF based on fBGD}
\begin{algorithmic}[1]
\label{al}
\STATE Initialize parameters randomly;
\WHILE {not converge}
\STATE Calculate intermediate tensors $M^{(\boldsymbol{\theta})},M^{(q)},M^{(u)},M^{(v)}$ based on Eq.(18)(20)(22)(23); [$O((n+m)K^2D+|X^+|(K+D))$]
\STATE Calculate intermediate vectors $S^{(\boldsymbol{\theta})},S^{(q)}$ based on Eq.(19)(21); [$O(n+m)D$]
\FOR{each user user $i$:}
\STATE Calculate gradients w.r.t $\boldsymbol{\theta}_i, \alpha_i, \boldsymbol{u}_i$ based on Eq.(1)(4)(5)(7-13)(24-27);  [$O((n+m)K^2D+|X^+|(K+D))$]
\ENDFOR
\FOR{each user item $j$:}
\STATE Calculate gradients w.r.t $w_j, b_j, \boldsymbol{v}_j$ based on Eq.(2)(3)(6)(7-13)(24-27);  [$O((n+m)K^2D+|X^+|(K+D))$]
\ENDFOR
\STATE update the parameters based on gradients.
\ENDWHILE
\end{algorithmic}
\end{algorithm}

\subsection{Evaluation Metrics}
 We adopt the following metrics to evaluate recommendation performance:
\begin{itemize}
\item Recall@K (Rec@K): This metric quantifies the fraction of consumed items that are in the top-K ranking list sorted by their estimated rankings. For each user $u$, we define $Rec(u)$ as the set of recommended items in top-K and $Con(u)$ as the set of consumed items in test data for user $u$. Then we have:
     \begin{small}
\begin{align}
    Recall@K&=\frac{1}{{|U|}}\sum\limits_{u \in U} {\frac{{|Rec(u)\cap Con(u)|}}{|Con(u)|}}
\end{align}
 \end{small}
\item Precision@K (Pre@K): This measures the fraction of the top-K items that are indeed consumed by the user:
 \begin{small}
 \begin{align}
    Precision@K&=\frac{1}{{|U|}}\sum\limits_{u \in U} {\frac{{|Rec(u)\cap Con(u)|}}{|Rec(u)|}}
\end{align}
 \end{small}
\item Normalized Discounted Cumulative Gain@K (NDCG@K): This is widely used in information retrieval and it measures the quality of ranking through discounted importance based on positions in the top-K recommendation lists. In recommendation, NDCG is computed as follow:
     \begin{small}
\begin{align}
NDCG@K &= \frac{1}{{|U|}}\sum\limits_{u \in U} {\frac{{DCG_u@K}}{{{IDCG_u@K}}}}
\end{align}
 \end{small}
where ${DC{G_u}}@K$ is defined as follow and ${{IDCG_u}}@K$ is the ideal value of ${{DCG_u}}@K$ coming from the best ranking.
 \begin{small}
\begin{align}
{DCG_u}@K &= \sum\limits_{i \in  Rec(u)\cap Con(u)} {\frac{1}{{{{\log }_2}(rank_{ui} + 1)}}}
\end{align}
\end{small}
where ${rank_{ui}}$ represents the rank of the item $i$ in the recommended list of the user $u$.
\item Mean Reciprocal Rank (MRR): Given the ranking lists, MRR is defined as follow:
\begin{align}
{MRR} &= \frac{1}{{|U|}}\sum\limits_{i \in U}{\sum\limits_{j \in  Con(i)} {\frac{1}{{rank_{ij}}}}}
\end{align}
MRR can be interpreted as the ease of finding all consumed items, as higher numbers indicate the consumed items are higher in the list.
\end{itemize}

\end{document}